\documentclass{pasj00}
\draft

\begin{document}
\SetRunningHead{H. Iwama, H. Asada, K. Yamada}
{Moment Approach to Astrometric Binary with Low SN} 
\Received{}%{yyyy/mm/dd}
\Accepted{}%{yyyy/mm/dd}

\title{Moment Approach for Determining 
the Orbital Elements of 
an Astrometric Binary 
with Low Signal-to-noise Ratio} 

%%% begin:list of authors
% Do NOT capitalize all letters in "textsc".
\author{Hirofumi \textsc{Iwama}, Hideki \textsc{Asada}, Kei \textsc{Yamada}} %
%  \thanks{Example: Present Address is xxxxxxxxxx}}
\affil{Faculty of Science and Technology, Hirosaki University, 
Hirosaki, Aomori 036-8561}
\email{asada@phys.hirosaki-u.ac.jp}

%%% end:list of authors

%%% Please use the following style in case that sorting by 
%%% affilation is impossible. 
%
% \author{%
%   D-Firstname \textsc{D-Familyname}\altaffilmark{1}
%   E-Firstname \textsc{E-Familyname}\altaffilmark{1,2}
%   and
%   F-Firstname \textsc{F-Familyname}\altaffilmark{2}}
% \altaffiltext{1}{Address of Institute}
% \email{ddddd@xxx.xxx.xx.xx}
% \email{eeeee@xxx.xxx.xx.xx}
% \altaffiltext{2}{Address of Institute}

%% `\KeyWords{}' always has to be placed before `\maketitle'.
 %Do NOT move this preamble from here!

\maketitle

\begin{abstract}
A moment approach for orbit determinations of 
an astrometric binary with low signal-to-noise ratio from 
astrometric observations alone 
is proposed, 
especially aiming at a close binary system 
with a short orbital period such as Cyg-X1 and  
also at 
a star wobbled by planets. 
As an exact solution to the 
nonlinearly coupled equation system, 
the orbital elements are written in terms of 
the second and third moments of 
projected positions that are measured by astrometry.
This may give a possible estimation of the true orbit.   
\end{abstract}

\section{Introduction}
Space astrometry missions such as 
Gaia and JASMINE are expected to reach a few micro arcseconds 
\citep{GAIA-1,GAIA-2,JASMINE}. 
Moreover, high-accuracy VLBI is also available. 

For visual binaries, formulations for orbit determinations 
have been well developed since the nineteenth century 
\citep{Thiele,Binnendijk,Aitken,Danby,Roy}. 
At present, numerical methods are successfully used 
\citep{EX,CO,OC}. 
Furthermore, an analytic solution for an astrometric binary, 
where one object can be observed and the other such as 
black holes and neutron stars is unseen, has been found 
\citep{AAK,Asada2008}. 
The solution requires that sufficiently accurate measurements 
of the position of a star (or the photo-center of a binary) 
are done more than four times for one orbital period of 
the binary system. 

For a close binary system with a short orbital period,  
we have a relatively large uncertainty in the position determination. 
For instance, the orbital period of Cyg-X1 is nearly 6 days, 
which are extremely shorter than that of normal binary stars, 
say a few months and several years. 
Because of such an extreme condition, 
it is interesting to seek another method 
in addition to the standard one. 
Moreover, stars with planets also are another interesting target. 

\begin{figure}
\includegraphics[width=160mm]{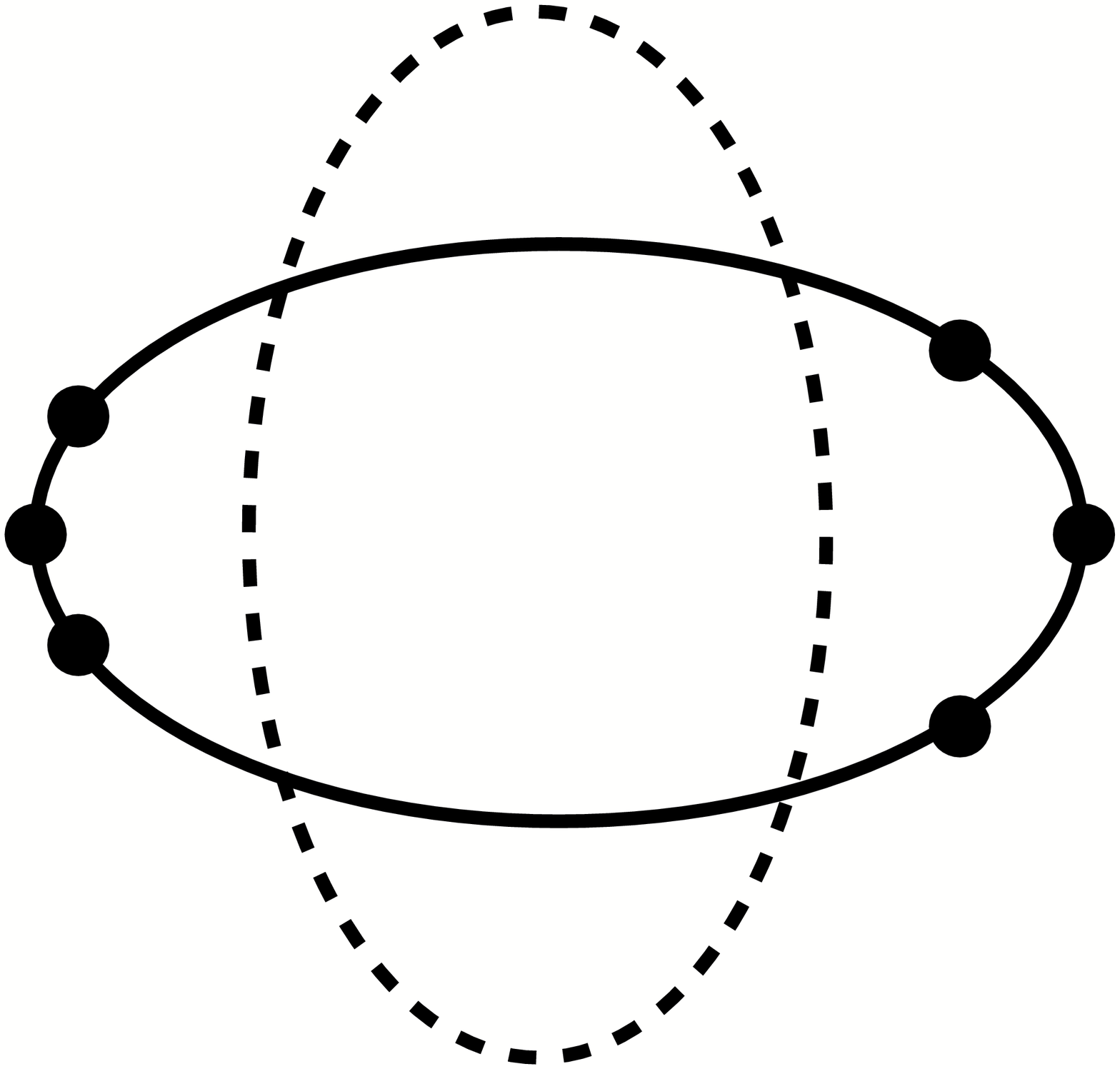}
\caption{
Comparison between the second moments 
$M_{xx}$ and $M_{yy}$. 
The orbit denoted by the closed solid curve 
has a larger variance along the $x$-axis, 
where $M_{xx}$ is larger than $M_{yy}$. 
On the other hand, 
for the orbit denoted by the closed dashed curve, 
$y$ components of the position have a larger scatter, 
where $M_{yy}$ is larger than $M_{xx}$. 
}
\label{moment-1}
\end{figure}

\begin{figure}
\includegraphics[width=160mm]{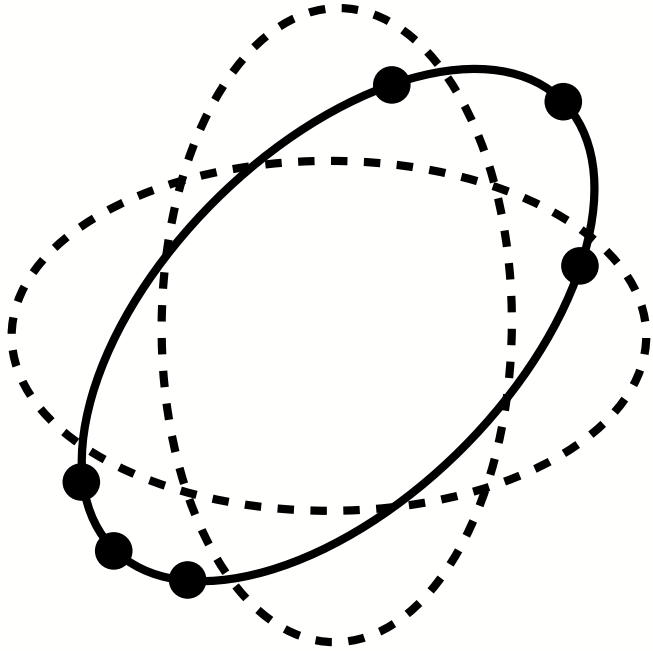}
\caption{
Dashed curves denote two orbits: 
One is the semimajor axis along the $x$-axis. 
The other is along the $y$-axis. 
The orbit denoted by the solid curve 
is not distinguished by using the second moments 
$M_{xx}$ and $M_{yy}$. 
The moment $M_{xy}$ is thus needed. 
}
\label{moment-2}
\end{figure}

What can we do for 
orbit determination from position measurements with 
low signal-to-noise (SN) ratio? 
It is expected that 
the position of the object is measured many times. 
The dense region of the observed points is 
corresponding to the neighborhood of the apastron of the Kepler orbit, 
because the motion of the source star is slower 
according to the Kepler's second law. 
On the other hand, a region of fewer points 
is including the periastron, around which 
the source star moves faster. 
Therefore, 
a statistical analysis including the variance of 
the measured positions and their correlation 
will bring the information about the orbital elements 
of the binary system. 
Gedanken experiments suggest that the second moments 
are useful for exploring the shape of the orbit 
but they are not sufficient for the full orbit determination.  
At least the third moments seem to be needed. 
See Figs. \ref{moment-1} and \ref{moment-2}. 

Therefore, the main purpose of this paper 
is to propose a method for orbit determination of 
an astrometric binary with low SN ratio 
by using the second and third moments. 
We shall provide also an exact solution 
for the coupled equations. 
As a result, the orbital elements of the binary 
are written in terms of the second and third moments.

This paper is organized as follows. 
We present a formulation and the solution in $\S$ 2. 
In $\S$ 3, numerical tests are also done to see how reliable 
the analytic result is for practical cases 
taking account of observational noises. 
$\S$ 4 is devoted to Conclusion.  
Throughout this paper, the spatial coordinates are the angular 
positions normalized by the distance to the celestial object.

\section{Basic Formulation} 
\subsection{Motion in the orbital plane}
Before projecting onto the celestial sphere, 
we consider a Kepler orbit, 
where the semimajor axis and semiminor one are denoted 
as $a_K$ and $b_K$. 
We choose the $(X, Y)$ coordinates on the orbital plane 
such that the $X$-axis is along the semimajor axis 
of the elliptic orbit, 
the $Y$-axis is along the semiminor one 
and the origin of the coordinates is chosen as the center 
of the ellipse. 
The orbit is expressed as  
\begin{eqnarray}
X &=& a_K \cos u , 
\\
Y &=& b_K \sin u , 
\end{eqnarray}
where $u$ denotes the eccentric anomaly. 

By introducing the eccentricity $e_K$, 
the semiminor axis becomes 
\begin{equation}
b_K = a_K \sqrt{1-e_K^2} .  
\end{equation}

For an object in the Kepler motion, 
the time $t$ is related with the position $u$ 
through the so-called Kepler equation 
\begin{equation}
t = t_0 + \frac{P}{2\pi} (u - e_K \sin u) ,
\label{Kepler}
\end{equation}
where $t_0$ denotes the time of periastron passage. 
This equation represents transcendental equations 
in the sense that it cannot be solved analytically 
without any approximation. 
In other words, $u$ cannot be expressed 
by using elementary functions of $t$. 
This makes the inverse problem in astrometry 
difficult.

\subsection{Projection with respect to the line of sight} 
We assume a plane perpendicular to the line of sight, 
where this plane is safely considered 
a small part of the celestial sphere 
because the angular position shift of 
extra-solar objects is sufficiently small. 
 
\begin{figure}
\includegraphics[width=100mm]{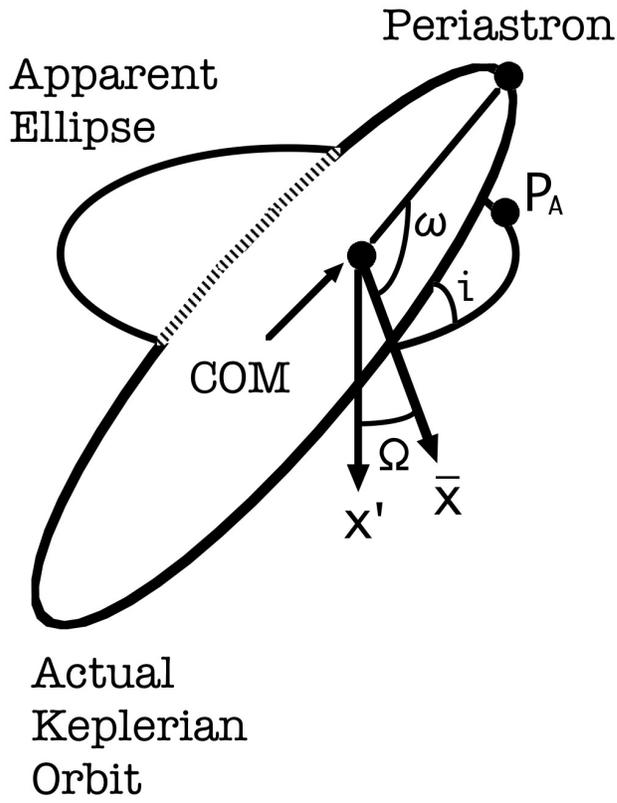}
\caption 
{Actual Keplerian orbit and apparent ellipse 
in three-dimensional space. 
We denote the inclination angle as $i$, 
the argument of periastron as $\omega$ and the longitude of 
ascending node as $\Omega$. 
These angles relate two coordinates $(x^{\prime}, y^{\prime})$ 
and $(\bar{x}, \bar{y})$, both of which choose the origin 
as the common center of mass. 
Here, the $x^{\prime}$ axis is taken to lie along the semimajor axis 
of the apparent ellipse, 
while the $\bar{x}$-axis is along 
the direction of the ascending node.
}
\label{config}
\end{figure}

A reference frame, which is used in astrometric observations, 
is described by the $(x, y)$ coordinates. 
As usual, the $(X, Y)$ coordinates on the orbital plane 
and the $(x, y)$ coordinates on the reference plane 
are related through 
\begin{eqnarray}
x &=& (X - a_K e_k) 
(\cos\omega\cos\Omega - \sin\omega\sin\Omega\cos i) 
\nonumber\\
&&
- Y (\sin\omega\cos\Omega + \cos\omega\sin\Omega\cos i)
\nonumber\\
&=&
x_0 + \alpha \cos u + \beta \sin u , 
\\
y &=& (X - a_K e_k) 
(\cos\omega\sin\Omega + \sin\omega\cos\Omega\cos i) 
\nonumber\\
&&
- Y (\sin\omega\sin\Omega - \cos\omega\cos\Omega\cos i)
\nonumber\\
&=&
y_0 + \gamma \cos u + \delta \sin u , 
\end{eqnarray}
where $x_0$, $y_0$, $\alpha$, $\beta$, $\gamma$, $\delta$ 
are defined as 
\begin{eqnarray}
x_0 &\equiv& 
- a_K e_K (\cos\omega\cos\Omega - \sin\omega\sin\Omega \cos i) , 
\label{x0}
\\
y_0 &\equiv&
- a_K e_K (\cos\omega\sin\Omega + \sin\omega\cos\Omega \cos i) , 
\label{y0}
\\
\alpha &\equiv&
a_K (\cos\omega\cos\Omega - \sin\omega\sin\Omega \cos i) , 
\label{alpha}
\\
\beta &\equiv&
-b_K (\sin\omega\cos\Omega + \cos\omega\sin\Omega \cos i) , 
\label{beta}
\\
\gamma &\equiv&
a_K (\cos\omega\sin\Omega + \sin\omega\cos\Omega \cos i) , 
\label{gamma}
\\
\delta &\equiv&
-b_K (\sin\omega\sin\Omega - \cos\omega\cos\Omega \cos i) .  
\label{delta}
\end{eqnarray}
Here, $\Omega$, $\omega$ and $i$ denote 
the longitude of ascending node, 
the argument of periastron 
and the inclination angle, respectively. 
See Fig. \ref{config}.

The ascending node and the descending one cannot be distinguished 
by astrometric observations alone. 
Therefore, both of $\pm$ for $i$ are possible. 
Furthermore, this paper focuses on the moments, 
so that the clockwise and anti-clockwise motions cannot be
distinguished. 
Two pairs of $(\Omega, \omega)$ are possible. 
Nevertheless, the shape of the orbit is uniquely determined 
as shown below.

\subsection{Moment Formalism}
Let us assume frequent observations of 
the angular position. 
Namely, we consider a large number of observed points. 
For such a case, 
the statistical average expressed as a summation 
is taken as the temporal average in an integral form as 
\begin{equation}
< F > \equiv \frac{1}{T_{obs}} \int_{0}^{T_{obs}} F dt , 
\label{average1}
\end{equation}
where $< \quad >$ denotes the mean and 
$T_{obs}$ denotes the total time duration of the observations.  

In this paper, we focus on the periodic motion, 
so that the above expression becomes 
the integration over one orbital period. 
We thus obtain 
\begin{eqnarray}
< F > &=& \frac{1}{P} \int_{t_0}^{t_0+P} F dt  
\nonumber\\
&=& 
\frac{1}{2\pi} \int_{0}^{2\pi} F (1 - e_K \cos u) du , 
\label{average2} 
\end{eqnarray}
where we used Eq. (\ref{Kepler}) and 
$dt = (1 - e_K \cos u) du$.

Let us consider statistical moments. 
Figure \ref{moment-1} suggests that the moments $M_{xx}$ and $M_{yy}$  
are useful for distinguishing two different orbits. 
Next, we consider the periastron for 
the orbit denoted by the closed solid curve. 
For simplicity, we assume that it 
is located at positive $x$ 
(in the right hand side of the ellipse). 
For this case, the object moves faster 
around the periastron and slower around the apastron. 
Hence, the dots in the figure schematically 
show asymmetry in the number of observed points. 
In order to distinguish such an asymmetry, 
the third moment such as $M_{xxx}$ seems useful because of 
the odd parity. 
Figure \ref{moment-2} suggests that $M_{xy}$ is needed 
via a Gedanken experiment. 
The second moments are defined as 
\begin{eqnarray}
M_{xx} &\equiv& 
< (x-<x>)^2 > 
\nonumber\\
&=& \frac12 (\alpha^2+\beta^2) - \frac14 e_K^2 \alpha^2 , 
\label{Mxx}
\\
M_{yy} &\equiv& 
< (y-<y>)^2 > 
\nonumber\\
&=& \frac12 (\gamma^2+\delta^2) - \frac14 e_K^2 \gamma^2 , 
\label{Myy}
\\
M_{xy} &\equiv& 
< (x-<x>) (y-<y>) > 
\nonumber\\
&=& \frac12 (\alpha\gamma+\beta\delta) - \frac14 e_K^2 \alpha\gamma , 
\label{Mxy}
\end{eqnarray}

The third moments are defined as 
\begin{eqnarray}
M_{xxx} &\equiv& 
< (x-<x>)^3 > 
\nonumber\\
&=& \frac38 e_K \alpha  (\alpha^2+\beta^2) 
- \frac14 e_K^3 \alpha^3 , 
\label{Mxxx}
\\
M_{yyy} &\equiv& 
< (y-<y>)^3 > 
\nonumber\\
&=& \frac38 e_K \gamma (\gamma^2+\delta^2) 
- \frac14 e_K^3 \gamma^3 , 
\label{Myyy}
\\
M_{xxy} &\equiv& 
< (x-<x>)^2 (y-<y>) > 
\nonumber\\
&=& \frac18 e_K (3\alpha^2\gamma+\beta^2\gamma+2\alpha\beta\delta) 
- \frac14 e_K^3 \alpha^2\gamma , 
\label{Mxxy}
\\
M_{xyy} &\equiv& 
< (x-<x>) (y-<y>)^2 > 
\nonumber\\
&=& \frac18 e_K (3\alpha\gamma^2+\alpha\delta^2+2\beta\gamma\delta) 
- \frac14 e_K^3 \alpha\gamma^2 . 
\label{Mxyy}
\end{eqnarray}
The moments $M_{xx}, \cdots, M_{xyy}$ are actually observables.

The vanishing of all the third moments leads 
to $e_K = 0$. 
In the following, let us consider $e_K \neq 0$ cases.

In the above second and third moments, 
the specific combinations among 
$\alpha$, $\beta$, $\gamma$, $\delta$ 
frequently appear. 
Therefore, it is convenient to define 
new variables as 
\begin{eqnarray}
I_1 &\equiv& \alpha^2+\beta^2 ,
\label{I-def}
\\
I_2 &\equiv& \gamma^2+\delta^2 ,
\label{II-def}
\\
I_3 &\equiv& \alpha\gamma+\beta\delta . 
\label{III-def}
\end{eqnarray}
By using the moments expressed by 
Eqs. (\ref{Mxx})-(\ref{Mxy}), 
the variables $I_1$, $I_2$, $I_3$ are rewritten as 
\begin{eqnarray}
I_1 &=& \frac12 e_K^2 \alpha^2 + 2 M_{xx} , 
\label{I}
\\
I_2 &=& \frac12 e_K^2 \gamma^2 + 2 M_{yy} ,
\label{II}
\\
I_3 &=& \frac12 e_K^2 \alpha\gamma + 2 M_{xy} .
\label{III}
\end{eqnarray}

Eq. (\ref{I}) is substituted into Eq. (\ref{Mxxx}) 
to obtain a cubic equation for 
$e_K \alpha$ as 
\begin{equation}
(e_K \alpha)^3 - 12 M_{xx} (e_K \alpha) + 16 M_{xxx} = 0 . 
\label{eKalpha}
\end{equation}
This cubic equation gives three roots as $e_K \alpha$ 
by using Cardano's formula. 
For saving the space, we do not write down the formula. 
In the similar manner, we obtain from 
Eqs. (\ref{II}) and (\ref{Myyy}) 
\begin{equation}
(e_K \gamma)^3 - 12 M_{yy} (e_K \gamma) + 16 M_{yyy} = 0 . 
\label{eKgamma}
\end{equation}
This provides $e_K \gamma$, where its multiplicity is three. 

At this point, we know both $e_K \alpha$ and $e_K \gamma$, 
which can be substituted into Eqs. (\ref{I})-(\ref{III}). 
We thus obtain the value of the variables 
$I_1$, $I_2$ and $I_3$.

In general, we get multi values for $e_K \alpha$ and $e_K \gamma$. 
However, 
they are true solutions for Eqs. (\ref{eKalpha}) and (\ref{eKgamma}), 
whereas not all of them satisfy the observed third moments. 
They have to satisfy 
the remaining set of the third moments as 
\begin{eqnarray}
M_{xxy} &=& 
\frac18 (e_K\gamma) I_1 + \frac14 (e_K \alpha) I_3 
- \frac14 (e_K \alpha)^2 (e_K\gamma) , 
\label{Eq-Mxxy}
\\
M_{xyy} &=& 
\frac18 (e_K\alpha) I_2 + \frac14 (e_K \gamma) I_3 
- \frac14 (e_K \alpha) (e_K\gamma)^2 , 
\label{Eq-Mxyy}
\end{eqnarray}
which can be used to pick up the correct $e_K \alpha$ 
and $e_K \gamma$ from multiple candidate values.

The definition of $I_1$ by Eq. (\ref{I-def}) is 
rewritten as  
\begin{equation}
\alpha^2 = -\beta^2 + I_1 ,
\end{equation}
which tells us $\alpha$ as a function of $\beta$ 
\begin{equation}
\alpha = \pm \sqrt{-\beta^2 + I_1} . 
\label{alphapm}
\end{equation}

It is obvious that 
the sign of the right hand side of Eq. (\ref{alphapm}) 
must be the same as that of the left hand side, namely $\alpha$. 
For $e_K \neq 0$, 
the sign of $\alpha$ is the same as that of $e_K \alpha$ 
that has been obtained above. 
This is expressed by 
\begin{equation}
\mbox{sgn}(\alpha) = \mbox{sgn}(e_K \alpha) , 
\end{equation}
where $\mbox{sgn}$ denotes the sign. 
Therefore, the sign of the right hand side of Eq. (\ref{alphapm}) 
is obtained uniquely as $\mbox{sgn}(e_K \alpha)$. 
Eq. (\ref{alphapm}) thus becomes  
\begin{equation}
\alpha = \mbox{sgn}(e_K \alpha) \sqrt{-\beta^2 + I_1} . 
\label{alphapm2}
\end{equation}
What is the difference between Eqs. (\ref{alphapm}) 
and (\ref{alphapm2})? 
Eq. (\ref{alphapm}) means two different equations 
because of $\pm$ in the right hand side, 
whereas 
Eq. (\ref{alphapm2}) is a {\it single} one.

For the later convenience, we define $\Gamma$ as 
\begin{equation}
\Gamma \equiv \frac{\gamma}{\alpha} , 
\label{Gamma}
\end{equation}
which is obtained from known quantities 
$e_K \alpha$ and $e_K \gamma$ as
$\Gamma = (e_K \gamma)(e_K \alpha)^{-1}$ . 

%\begin{equation}
%\Gamma = \frac{e_K \gamma}{e_K \alpha} . 
%\label{Gamma2}
%\end{equation}

We substitute Eq. (\ref{Gamma}) into 
the definition of $I_3$ by Eq. (\ref{III-def})  
in order to delete $\gamma$.  
We obtain 
\begin{equation}
\alpha^2 = \frac{I_3 - \beta\delta}{\Gamma} . 
\end{equation}
This is substituted into the definition of $I_1$ 
by Eq. (\ref{I-def}). 
We obtain  
\begin{equation}
\delta = \frac{\beta^2 \Gamma + I_3 - I_1 \Gamma}
{\beta} . 
\label{delta-fn}
\end{equation}
This is a function of only $\beta$.

Eqs. (\ref{Gamma}) and (\ref{delta-fn}) 
are substituted into $\gamma$ and $\delta$ in Eq. (\ref{II-def}). 
After rather lengthy calculations, we get 
\begin{equation}
\beta^2 = \frac{(I_3 - I_1 \Gamma)^2}
{I_1 \Gamma^2 + I_2 - 2 I_3 \Gamma} ,
\label{beta^2}
\end{equation}
where we used Eq. (\ref{I-def}) for $\alpha^2$. 
Interestingly, the right hand side of this equation 
consists of only the known quantities $I_1$, $I_2$, 
$I_3$ and $\Gamma$. 
Therefore, we obtain the value of $\beta^2$, 
which determines $\alpha$ through Eq. (\ref{alphapm2}). 

{}Eq. (\ref{beta^2}) is solved for $\beta$ as 
\begin{equation}
\beta = \pm \frac{I_3 - I_1 \Gamma}
{\sqrt{I_1 \Gamma^2 + I_2 - 2 I_3 \Gamma}} . 
\label{beta-sol}
\end{equation}
Unfortunately, we do not know $e_K \beta$ 
contrary to $e_K \alpha$. 
Therefore, the sign of the right hand side of Eq. (\ref{beta-sol}) 
is not uniquely determined. 
The multiplicity of $\beta$ is two and 
hence that of $\delta$ is also two according to Eq. (\ref{delta-fn}). 

Up to this point, we know the value of $\alpha$ and $e_K \alpha$ from 
observed quantities. 
We thus find separately $e_K$ as 
$e_K = (e_K \alpha)\alpha^{-1}$ . 
%\begin{equation}
%e_K = \frac{e_K \alpha}{\alpha} . 
%\end{equation}
It is crucial in the following procedure 
that the eccentricity $e_K$ is determined at this step.

It is very inconvenient that 
$\beta$ and $\delta$ are proportional to 
the semiminor axis $b_K$ in their definition. 
We know $e_K$, so that $b_K$ can be expressed by $a_K$. 
Hence, we define renormalized quantities as 
\begin{eqnarray}
\tilde\beta &\equiv&
\frac{\beta}{\sqrt{1-e_K^2}} \nonumber\\
&=&
-a_K (\sin\omega\cos\Omega + \cos\omega\sin\Omega \cos i) , \\
\tilde\delta &\equiv&
\frac{\delta}{\sqrt{1-e_K^2}} \nonumber\\
&=&
-a_K (\sin\omega\sin\Omega - \cos\omega\cos\Omega \cos i) ,
\end{eqnarray}
where we used Eqs. (\ref{beta}) and (\ref{delta}). 
We know the values of $\beta$, $\delta$ and $e_K$. 
Therefore, one can estimate $\tilde\beta$ and $\tilde\delta$.

One can construct from four variables 
$\alpha$, $\tilde\beta$, $\gamma$, $\tilde\delta$,  
some quantities 
that are dependent on the inclination angle $i$ 
but not on any other angles $\omega$ nor $\Omega$. 
One example is 
\begin{eqnarray}
C &\equiv& \alpha^2 + \tilde\beta^2 + \gamma^2 + \tilde\delta^2
\nonumber\\
&=& a_K^2 (1+\cos^2 i) .
\label{C}
\end{eqnarray}
Another is 
\begin{eqnarray}
D &\equiv& \alpha \tilde\delta - \tilde\beta \gamma
\nonumber\\
&=& a_K^2 \cos i . 
\label{D}
\end{eqnarray}
These relations can be verified by direct calculations. 
Note that $D$ must be positive because $\cos i \geq 0$. 
This positivity chooses one pair of $(\tilde\beta, \tilde\delta)$ 
and reject the other pair. 
Only the pair of $(\beta, \delta)$ is thus obtained. 
We know $\alpha$, $\gamma$, $\tilde\beta$, $\tilde\delta$, 
so that $C$ and $D$ can be estimated. 

By deleting $a_K^2$ from Eqs. (\ref{C}) and (\ref{D}), 
we obtain  
\begin{equation}
\cos^2 i - \frac{C}{D} \cos i + 1 = 0 .
\label{cos2i}
\end{equation}
This is a quadratic equation for $\cos i$. 
Apparently, two cases of $\cos i$ are possible. 
However, this is not the case. 
By using Newton's identities for Eq. (\ref{cos2i}), 
we obtain the identity as 
\begin{equation}
(\cos i_1) \times (\cos i_2) = 1 , 
\end{equation}
where $\cos i_1$ and $\cos i_2$ denote 
the two roots for Eq. (\ref{cos2i}). 
The inequality as 
\begin{equation}
|\cos i| \leq 1 , 
\end{equation}
thus leads to the unique value of $\cos i$, 
because the other exceeds the unity as its absolute value. 

The single value of $\cos i$ provides 
positive $i$ and negative one. 
They are corresponding to the ascending node and 
the descending node, respectively. 
They cannot be distinguished by astrometric observations alone. 
In order to distinguish them, radial velocity 
measurements for instance are needed.

We get the value of the inclination angle. 
{}From Eq. (\ref{D}), therefore, 
the semimajor axis is obtained as 
\begin{equation}
a_K = \sqrt{\frac{D}{\cos i}} . 
\end{equation}
This suggests that the value of $a_K$ is 
uniquely determined but not doubly.

In order to determine $\omega$ and $\Omega$, 
let us consider other combinations among 
$\alpha$, $\gamma$, $\tilde\beta$, $\tilde\delta$. 

Direct calculations lead to 
\begin{eqnarray}
\alpha^2 + \tilde\beta^2
= a_K^2 (\cos^2\Omega + \sin^2\Omega \cos^2 i) , \\
\gamma^2 + \tilde\delta^2
= a_K^2 (\sin^2\Omega + \cos^2\Omega \cos^2 i) . 
\end{eqnarray}
The ratio of them is denoted as 
\begin{eqnarray}
r_1 &\equiv&
\frac{\gamma^2 + \tilde\delta^2}{\alpha^2 + \tilde\beta^2} 
\nonumber\\
&=& 
\frac{\sin^2\Omega + \cos^2\Omega \cos^2 i}
{\cos^2\Omega + \sin^2\Omega \cos^2 i} . 
\label{r1}
\end{eqnarray}
This is solved for $\Omega$ as 
\begin{equation}
\tan^2\Omega
= \frac{r_1 - \cos^2 i}{1 - r_1 \cos^2 i} ,
\end{equation}
which gives the values of $\Omega$ 
because we have already determined $i$ and $r_1$. 
As mentioned already, both of $\pm\Omega$ are allowed.

Next, we consider different combinations as 
\begin{eqnarray}
\alpha^2 + \gamma^2
= a_K^2 (\cos^2\omega + \sin^2\omega \cos^2 i) , \\
\tilde\beta^2 + \tilde\delta^2
= a_K^2 (\sin^2\omega + \cos^2\omega \cos^2 i) ,
\end{eqnarray}
which can be verified by direct calculations. 
The ratio of them is denoted as 
\begin{eqnarray}
r_2 &\equiv&
\frac{\tilde\beta^2 + \tilde\delta^2}{\alpha^2 + \gamma^2} 
\nonumber\\
&=& \frac{\sin^2\omega + \cos^2\omega \cos^2 i}
{\cos^2\omega + \sin^2\omega \cos^2 i} .
\label{r2} 
\end{eqnarray}
This is solved for $\omega$ as 
\begin{equation}
\tan^2\omega
= \frac{r_2 - \cos^2 i}{1 - r_2 \cos^2 i} ,
\end{equation}
which tells us the values of $\omega$ 
because we have already determined $i$ and $r_2$. 
As mentioned already, both of $\pm\omega$ are possible.

At most four values of $\Omega$ are possible. 
Similarly, the maximum multiplicity of $\omega$ is four. 
In total, sixteen sets of $(\omega, \Omega)$ appear to exist. 
The multiplicity of $(\omega, \Omega)$ is reduced, because 
they must satisfy the definition of $\alpha$, 
$\beta$, $\gamma$, $\delta$ with the uniquely determined 
$a_K$, $e_K$ and $\cos i$. 
In particular, $\alpha$, $\beta$, $\gamma$, $\delta$ 
include different combinations of $\sin$ and $\cos$. 
Basically, the sign of $\sin$ and $\cos$ has four types 
$(+, +)$, $(+, -)$, $(-, +)$, $(-, -)$. 
Hence, the apparent multiplicity sixteen is divided by four. 
In addition, it is obvious from Eqs. (\ref{alpha})-(\ref{delta}) 
that the sign of $\Omega$ depends on that of $\omega$. 
The multiplicity is thus divided also by two. 
As a consequence, 
we get only the two pairs of $(\Omega, \omega)$. 
One pair is corresponding to the clockwise motion 
and the other to the anti-clockwise one.

\section{Discussion}
\subsection{Numerical test}
The above formalism is discussed in an idealized world. 
Numerical tests are done below to see whether 
the analytic result works for practical cases. 
First, Eqs. (\ref{average1}) and (\ref{average2}) 
assume that one can integrate observed quantities. 
In practice, however, observations are discrete, 
for which the integration becomes a summation. 
The integration and the summation could agree 
in the limit that the number of observations $N$ 
approaches the infinity. 
In a real world, $N$ is a large number but much smaller 
than the infinity. 
Does the above formalism still give a reliable answer? 
For investigating this point, 
we perform numerical simulations. 
According to the simulation for observations with equal time interval, 
the above formalism recovers perfectly 
the orbital parameters for $N = 100$. 

Next, we consider observation noises. 
The above formalism assumes that the observed points 
are located on an apparent ellipse. 
However, position measurements are inevitably 
associated with observational errors. 
Therefore, we perform numerical simulations by adding 
Gaussian 
errors into the position measurements 
as 
$x \to x +\Delta x$ and $y \to y +\Delta y$, 
where $\Delta x$ and $\Delta y$ obey 
Gaussian distributions with the standard deviation $\sigma$. 
We consider two cases: 
$\sigma = 0.1$ (smaller case) and 
$0.5$ (larger one) 
in the units of $a_K=1$. 
Table $\ref{table}$ is a list of the orbital parameters that are 
recovered by using the above formalism. 
See Fig. $\ref{points}$ for simulated points 
in a $N=100$ simulation for $\sigma = 0.1$ and $0.5$. 
For small observation error cases, the orbital parameters are 
well recovered. 
For large observation errors comparable to a half of 
the semimajor axis, however, the recovered angles 
$i$, $\omega$ and $\Omega$ are far from the true ones. 
On the other hand, the eccentricity $e_K$ and 
the semimajor axis $a_K$ are recovered better than the angles. 
The semimajor axis is overestimated, 
because the simulated second moments apparently become larger than 
the true ones owing to such a large dispersion. 
We numerically study also different parameter values. 
They lead to similar results, 
for which numerical tables are omitted for saving the space. 

Our numerical tests for the discreteness of observations 
and for the observation noises suggest that 
the above formalism derived for the idealized system 
could work in practice if observation noises are not so large. 

The previous analytical method cannot guess 
orbit parameters for such large measurement errors 
$\sigma = 0.5$, mostly because $\cos i$ apparently 
becomes larger than the unity 
(namely, the data points fit better with open orbits) \citep{AAK2}. 
 
Up to this point, we have assumed that the orbital period is known. 
What happens for unknown binaries? 
We have made Fourier analyses for 
numerically simulated points with time in Fig. \ref{points}. 
In the time domain, the Fourier spectrum has two peaks. 
One peak corresponds to the orbital period and the other is 
around the artificial time step in the numerical simulations. 
This suggests that the moment approach can be applied also to 
unknown binary systems, if a Fourier analysis is adequately used 
to know the orbital period.  
Namely, the present method could be used to search new binary systems.

Let us consider Cyg-X1, $a \sim 0.2$ AU at $2$ kpc from us. 
The expected angular accuracy in JASMINE  
is $\sim 10$ microarcsec, so that 
the semimajor axis of Cyg-X1 can become a direct observable. 
Other known X-ray binaries seem too faint to 
be observed by JASMINE.

A Sun-like star at 20 pc with Jupiter-like planet at 1 AU 
could produce a wobble of 0.001 AU, corresponding to 50 microarcsec., 
which must be an interesting target.

\subsection{Proper motion of the binary}
In the main part of this paper, 
we ignore the proper motion 
of the binary system in our galaxy. 
This is mostly because, 
for a close binary, the proper motion of the binary 
in our galaxy causes a larger cumulative displacement 
than the orbital motion of the component stars, 
though the orbital velocity may be larger than the proper motion. 
In advance, therefore, we know the proper motion  
before determining the orbital elements. 
For instance, it can be done by a comparison 
between the Hipparcos data and the future space astrometry. 
If one wishes to determine the proper motion $(v_x, v_y)$ 
in the present formalism, however, 
the apparent positions should be replaced as 
$x \to x + v_x t$ and $y \to y + v_y t$. 
By averaging the observed position according to Eq. (\ref{average1}), 
we obtain 
\begin{eqnarray}
<x> &=& \frac12 T_{obs} v_x + const. 
+ O\left( \frac{T_K}{T_{obs}} a_K \right) , 
\nonumber\\
<y> &=& \frac12 T_{obs} v_y + const. 
+ O\left( \frac{T_K}{T_{obs}} a_K \right) , 
\end{eqnarray}
where 
the terms of $O(a_K T_K T_{obs}^{-1})$ 
in the right hand side 
come from the Kepler motion. 
Hence the terms are nothing but 
a decaying term such as $T_{obs}^{-1} \cos u_{obs}$ 
for the eccentric anomaly $u_{obs}$ 
corresponding to $t=T_{obs}$. 
They become negligible 
as $T_{obs} T_K^{-1} \to \infty$. 
Furthermore, the cumulative translation  
by the proper motion exceeds the oscillatory displacement 
by the Kepler motion for a close binary. 
That is, $a_K (T_{obs} v)^{-1} \ll 1$, 
where $v \equiv \sqrt{v_x^2 + v_y^2}$. 
Owing to this effect, 
the decaying part due to the Kepler motion 
goes away for a long observation period $T_{obs}$, 
say $> 1$ yr.

The parts growing linearly in the observation time $T_{obs}$ 
give the information about $(v_x, v_y)$, 
provided $T_{obs}$ is taken as a variable in the data analysis.   
If $T_{obs} \gg T_K$, 
the above extraction of the linearly growing part 
will be possible, especially for a short-separation binary. 

% \subsection{Numerical Tests}

\subsection{Mildly relativistic compact binary} 
Finally, we mention a mildly relativistic compact binary, 
in which the relativistic advance of the periastron occurs. 
Rigorously speaking, we have to take account of the general 
relativistic equations. 
In practice, however, the above method may be applied. 
For instance, let us imagine two-year observations. 
For the first-year data, the periastron direction $\varpi_1$ 
is derived. 
The second-year data tells $\varpi_2$. 
The difference between the first-year and second-year directions 
of the periastron suggests the periastron advance as 
$(\varpi_2-\varpi_1) \mbox{yr}^{-1}$. 

For instance, the Hulse-Taylor binary pulsar 
shows the large periastron shift rate $\dot\varpi = 4$ 
deg. per year, 
though the angular orbital radius is a few microarcsec. 
and it is beyond the current measurement capability.

\section{Conclusion}
This paper proposed 
a moment approach for 
orbit determinations of a close binary system 
with a short orbital period from astrometric observations alone. 
As an exact solution to the coupled equations, 
the orbital elements are written in terms of 
the second and third moments of the 
projected position that is measured by astrometry. 

The moment formalism does not replace 
the standard method 
using Kepler equation. 
It is safer to say that the present formalism 
is a supplementary tool for giving a rough 
parameter estimation, 
which can be used as a trial value 
for full numerical data fittings. 
It is interesting to make numerical tests of the present method. 
It is left as a future work. 

\begin{figure}
\includegraphics[width=140mm]{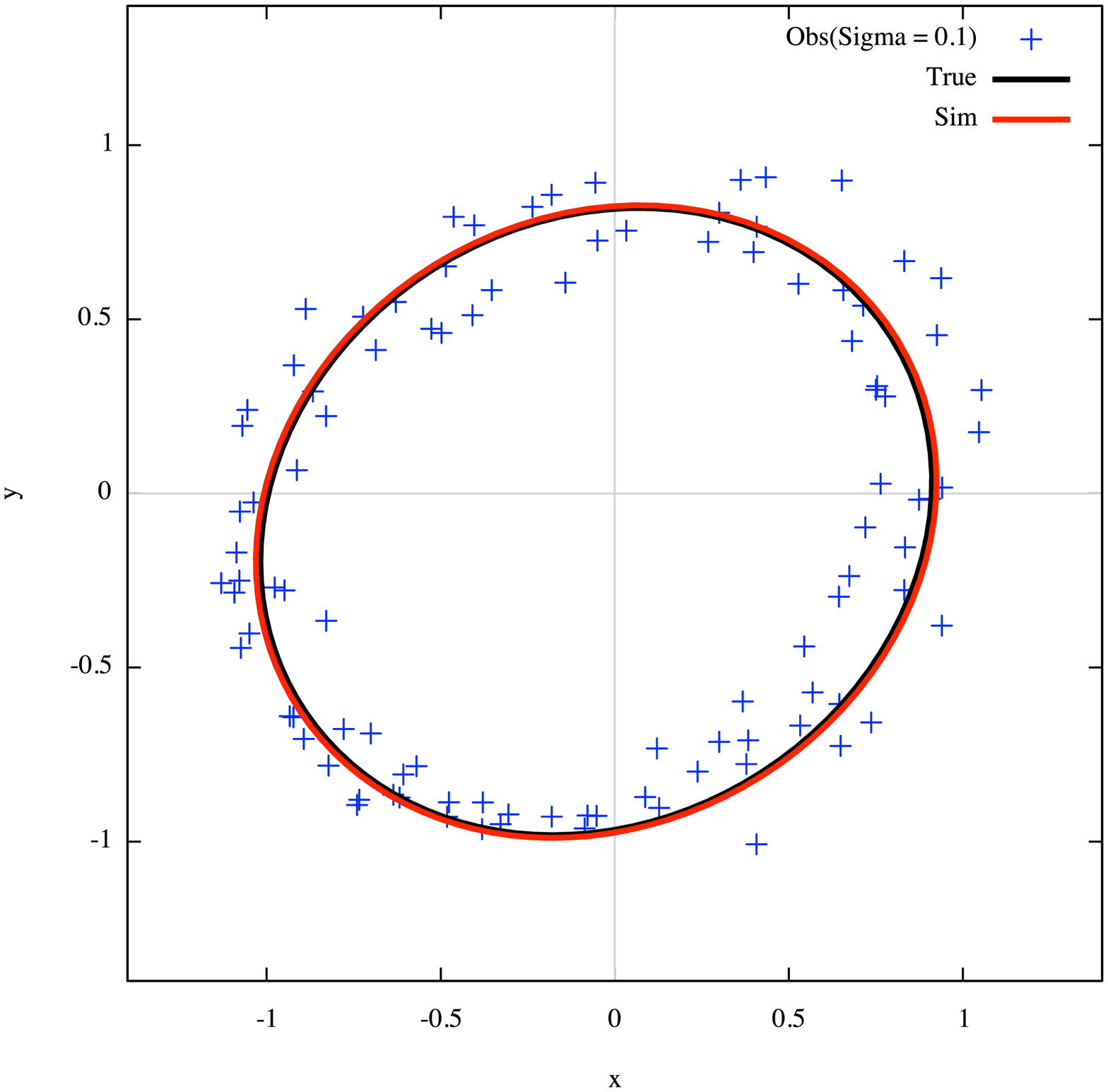}
\includegraphics[width=140mm]{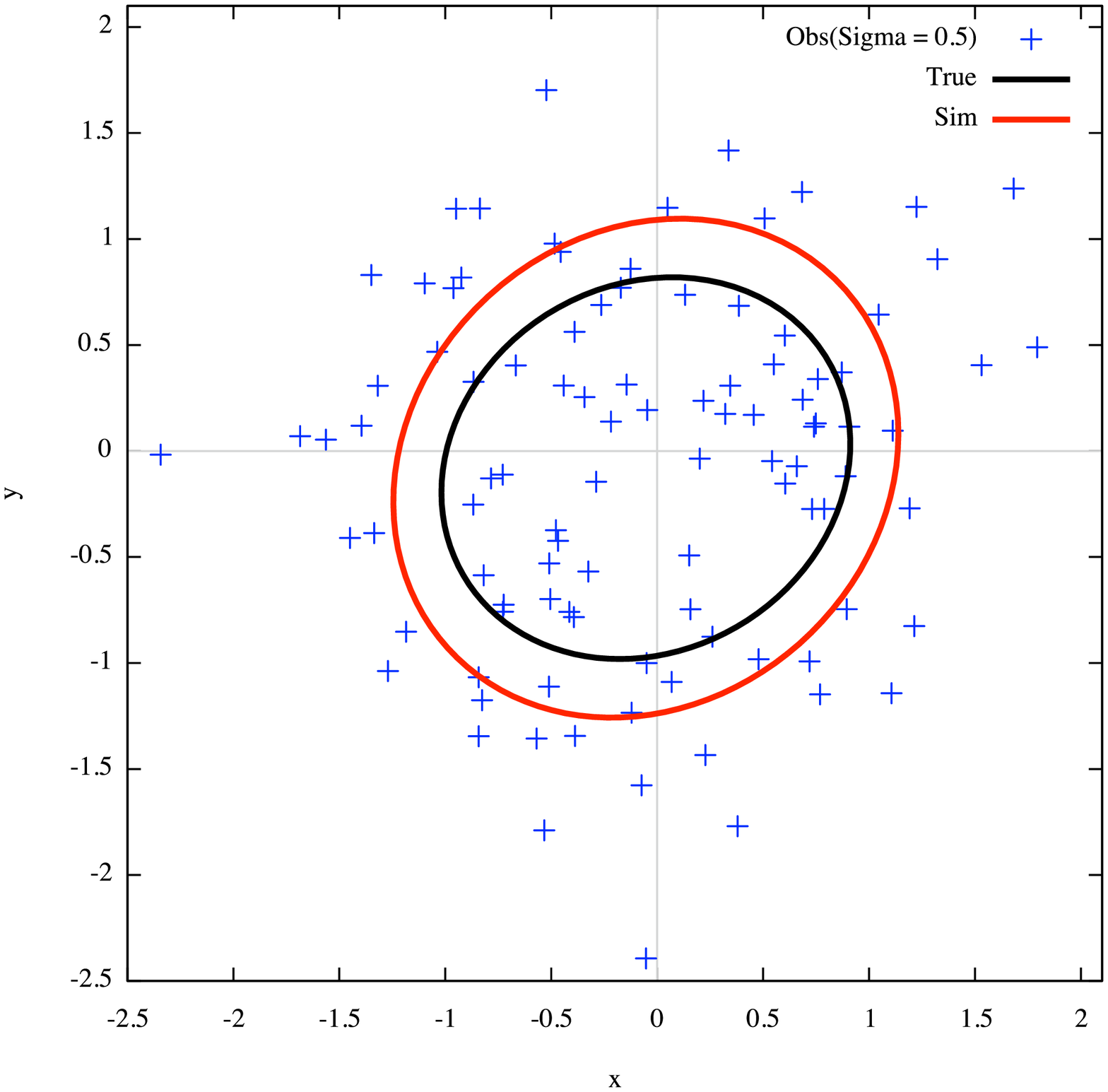}
\caption{
One hundred observed points 
of the same source star in the $x-y$ plane 
for a simulation  
with constant time interval. 
The parameters are $\omega = 30$ [deg.], 
$\Omega = 30$ [deg.], $i = 30$ [deg.], 
$e_K = 0.1$, $a_K = 1.0$ and $N=100$.   
The black curve and the gray (red in color) one denote orbits 
for the true parameter and for the mean value 
of the recovered parameters, respectively. 
Top: 
Gaussian errors $\sigma = 0.1$. 
Bottom: $\sigma = 0.5$. 
}
\label{points}
\end{figure}

In the moment approach, 
the temporal information is smeared. 
Therefore, the orbital period cannot be determined 
by this method. 
Another method such as the Fourier analysis of 
position data with time (in Fig. \ref{points} for example) 
would give a characteristic frequency that is 
the inverse of the orbital period.  
Fourier analyses recover the orbital period from 
numerically simulated data for Fig. \ref{points}. 
This suggests that the moment approach can be applied also to 
unknown binary systems, if a Fourier analysis is adequately used to know 
the orbital period.  
Namely, the method could be used to search new binary systems.

%\acknowledgments
We would like to thank Professor N. Gouda for useful information 
on astrometry missions. 
We wish to thank the JASMINE science WG member 
for stimulating conversations. 
We would be grateful to Y. Sendouda, R. Takahashi and K. Izumi 
for helpful conversations on numerical simulations. 
This work was supported in part (H.A.) 
by a Japanese Grant-in-Aid 
for Scientific Research from the Ministry of Education, 
No. 21540252 (Kiban-C) 
and in part (K.Y.) by JSPS research fellowship for young scientists.

%%%
% See the manual for the detail.
%%%

\clearpage

\clearpage

\clearpage

\clearpage

\begin{figure}
\includegraphics[width=100mm]{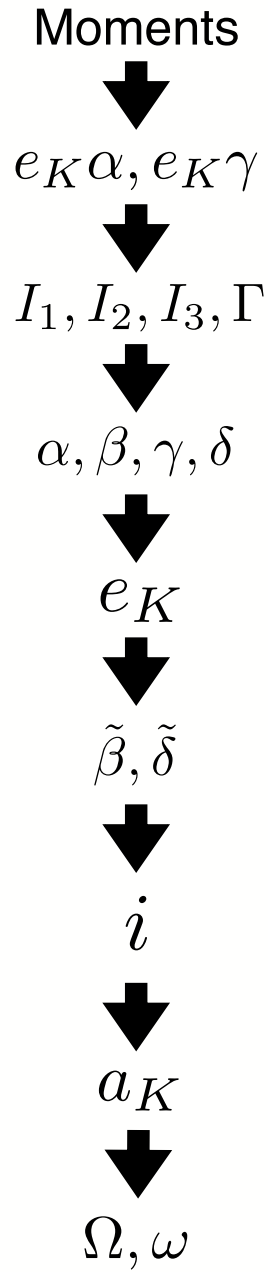}
\caption{
Flowchart of parameter determinations in the moment formalism. 
The starting point is 
evaluating the moments from astrometric observations. 
}
\label{flowchart}
\end{figure}

\clearpage

\clearpage

%% If you use the table environment, please indicate horizontal rules using
%% \tableline, not \hline.
%% Do not put multiple tabular environments within a single table.
%% The optional \label should appear inside the \caption command.

%% If the table is more than one page long, the width of the table can vary
%% from page to page when the default \tablewidth is used, as below.  The
%% individual table widths for each page will be written to the log file; a
%% maximum tablewidth for the table can be computed from these values.
%% The \tablewidth argument can then be reset and the file reprocessed, so
%% that the table is of uniform width throughout. Try getting the widths
%% from the log file and changing the \tablewidth parameter to see how
%% adjusting this value affects table formatting.

%% The \dataset{} macro has also been applied to a few of the objects to
%% show how many observations can be tagged in a table.

%% Tables may also be prepared as separate files. See the accompanying
%% sample file table.tex for an example of an external table file.
%% To include an external file in your main document, use the \input
%% command. Uncomment the line below to include table.tex in this
%% sample file. (Note that you will need to comment out the \documentclass,
%% \begin{document}, and \end{document} commands from table.tex if you want
%% to include it in this document.)

%% \input{table}

%% The following command ends your manuscript. LaTeX will ignore any text
%% that appears after it.
\clearpage

%\begin{deluxetable}{ccrrrrrrrrcrl}
%\begin{deluxetable}{crrrrr}
%\tabletypesize{\scriptsize}
%\rotate
%\tablecaption{
\begin{table}
\caption{
Reconstructing the parameters for numerical simulations 
for three different eccentricity cases as $e_K = 0.1, 0.3, 0.5$. 
In the table, the row $\sigma = 0$ indicates true orbital parameters, 
whereas the rows $\sigma = 0.1$ and $0.5$  
provide the recovered values for adding Gaussian errors 
(0.1 or 0.5 in the units of the true semimajor axis, respectively). 
For each parameter set, 100 runs are done and 
the mean and the standard deviation are also evaluated. 
} 
%\tablewidth{0pt}
%\tablehead{
%\colhead{ Type } & \colhead{$\omega$ [deg.]} & 
%\colhead{$i$ [deg.]} & \colhead{$\Omega$ [deg.]} & 
%\colhead{$e_K$} & \colhead{$a_K$}
%}
%\startdata
%$Error=1/1000$%
%\arrayrulewidth=0.2mm
%\doublerulesep=0.15mm
%\begin{table}[h!]
%\begin{table}[p]
%\caption{Reconstructing the parameters for small observational errors}
%\renewcommand{\arraystretch}{0.5}
%$$ \begin{array}{|l||r|r|r|r|}\hline
%a_k-e_k-i-\omega&{\Delta{a}}&\hspace{10mm}{\Delta{e}}&
%\hspace{10mm}{\Delta{i}}&\hspace{10mm}{\Delta{\omega}}\\
%\hline\hline\hline\hline
\begin{center}
\begin{tabular}{l|lllll}
$\sigma$ & $e_k$ & $a_k$ & $i$ [deg.] & $\omega$ [deg.] & $\Omega$ [deg.] \\ 
\hline
0 & $0.1$ & $1.0$ & $30.0$ & $30.0$ & $30.0$ \\
0.1 & $0.09454 \pm 0.02518$ & $1.011 \pm 0.01423$ 
& $30.02 \pm 2.039$ & $22.37 \pm 17.45$ & $29.81 \pm 4.205$ \\
0.5 & $0.1445 \pm 0.09343$ & $1.257 \pm 0.07624$ 
& $28.87 \pm 7.731$ & $43.87 \pm 25.79$ & $40.75 \pm 22.62$ \\
\hline
0 & $0.3$ & $1.0$ & $30.0$ & $30.0$ & $30.0$ \\
0.1 & $0.2936 \pm 0.02269$ & $1.011 \pm 0.01526$ 
& $30.04 \pm 2.228$ & $28.73 \pm 8.557$ & $30.20 \pm 4.687$ \\
0.5 & $0.2327 \pm 0.1211$ & $1.266 \pm 0.08215$ 
& $29.63 \pm 7.189$ & $41.26 \pm 25.13$ & $35.45 \pm 18.14$ \\
\hline
0 & $0.5$ & $1.0$ & $30.0$ & $30.0$ & $30.0$ \\
0.1 & $0.4793 \pm 0.03103$ & $1.005 \pm 0.02253$ 
& $30.34 \pm 2.477$ & $27.53 \pm 8.234$ & $31.72 \pm 5.709$ \\
0.5 & $0.3065 \pm 0.1223$ & $1.233 \pm 0.07314$ 
& $31.33 \pm 5.519$ & $39.78 \pm 26.46$ & $40.78 \pm 16.08$ \\

%\enddata
%\hline
%\end{array} $$

%\end{table}

\label{table}
%\end{deluxetable}
\end{tabular}
\end{center}
\end{table}

\end{document}